# Bolometric and non-bolometric radio frequency detection in a metallic single-walled carbon nanotube

Daniel F. Santavicca,[1] Joel D. Chudow,[1] Daniel E. Prober,[1*] Meninder S. Purewal,[2] and Philip Kim[2]

[1]Department of Applied Physics, Yale University

[2]Departments of Physics and Applied Physics, Columbia University

*e-mail: daniel.prober@yale.edu

**Abstract**

We characterize radio frequency detection in a high-quality metallic single-walled carbon nanotube. At a bath temperature of 77 K, only bolometric (thermal) detection is seen. At a bath temperature of 4.2 K and low bias current, the response is due instead to the electrical nonlinearity of the non-ohmic contacts. At higher bias currents, the contacts recover ohmic behavior and the observed response agrees well with the calculated bolometric responsivity. The bolometric response is expected to operate at terahertz frequencies, and we discuss some of the practical issues associated with developing high frequency detectors based on carbon nanotubes.

Carbon nanotubes have been studied for a number of detector applications, including microwave,[1,2] terahertz (THz),[3,4] and infrared[5,6] detection. The extremely small specific heat of a carbon nanotube is predicted to give a bolometric (thermal) detector with a very fast response time and good sensitivity.[7] A power detector with a sufficiently fast response can be used as a heterodyne mixer to detect the power envelope of the combined signal and local oscillator (LO). Its output oscillates at the intermediate (difference) frequency (IF). THz heterodyne detectors based on superconducting bolometers, superconducting tunnel junctions, and Schottky diodes have found important applications in radioastronomy[8,9] and laboratory spectroscopy.[10] Superconducting detectors are more sensitive and require smaller LO power, but Schottky diodes have a larger IF bandwidth and can operate at higher temperatures. Carbon nanotubes offer the potential for a THz heterodyne detector with a very large IF bandwidth and modest LO power requirements,[7] and hence may prove to be an attractive complement to superconducting detectors and Schottky diodes.

We study the response of an individual metallic carbon nanotube with high quality palladium contacts. The heterodyne response can be due to two mechanisms: (1) bolometric detection due to heating of a device with a temperature-dependent resistance, or (2) a nonlinear current-voltage (*I-V*) characteristic that is not thermal in origin. Previous work studied individual[2] and bundles of[3,4] carbon nanotubes. These samples had very high contact resistance, ~M$\Omega$ per nanotube. This produced at lower frequencies (<< 1 THz) a large non-thermal mixer response due to the contacts' non-linear *I-V* characteristic. Fu *et al*. showed that the parallel capacitance associated with high contact resistance decreases the effect of the contact non-linearity at THz frequencies.[4] In lower frequency studies, this contact response can be large, and can mask the bolometric response.[2,3] Macroscopic mats of suspended carbon nanotubes have also been investigated as power detectors,[5,6] but the sensitivity of these devices is modest and the thermal response is very slow, ~ms. The studies of nanotube bundles identified the bolometric mechanism as being responsible for THz detection.[4] Due to uncertainties in the properties of the nanotube bundles, however, quantitative comparisons to theory were difficult. The goals of the present work are to study radio frequency detection over a range of temperature and bias currents, to identify the bolometric contribution useable at THz frequencies, and to provide a quantitative comparison to theory.

To fabricate devices, carbon nanotubes are grown using chemical vapor deposition on degenerately doped silicon with a 500 nm thick oxide ($SiO_2$).[11] Measurements of the nanotube height (2.0 ± 0.2 nm) and the saturation current confirm that it is a single-walled nanotube.[12] Deposited palladium contacts are used to achieve low contact resistance. The silicon substrate is used as a global back-gate. The nanotube displays a decrease in conductance near zero gate voltage, which is attributed to a small curvature-induced bandgap. All data reported here were taken at a gate voltage of -30 V. For gate voltages below ≈ -20 V, the conductance is large and is insensitive to small changes in the gate potential. By measuring the resistance of segments of different lengths from a single physical nanotube, we infer a temperature-independent contact resistance of 8 ± 1 kΩ,[12] close to the ideal contact resistance of $h/(4e^2) \approx 6.4$ kΩ for four ballistic quantum channels. The additional internal resistance of each nanotube segment is ≈ 1 kΩ/μm at 4 K, and increases to ≈ 2 kΩ/μm at 77 K. Detailed dc characteristics of this same sample have been reported previously in ref. 12. While nanotube segments of several different lengths have been measured, we focus here on the results from the 5 μm length. This is sufficiently long to exhibit diffusive transport at 77 K, and also for larger currents at 4.2 K. Measurements of longer samples have a reduced signal-to-noise ratio due to the increased impedance mismatch between the nanotube and the amplifier, but are consistent with the results reported here.

In figure 1 we plot the dc resistance $R = V/I$ measured at low current ($I = 100$ nA) as a function of the bath temperature $T_b$. We also plot the dc resistance as a function of the bias current $R(I)$ at $T_b = 4.2$ K and 77 K. The increase in resistance with increasing bias current at 77 K, and for $|I| > 0.4$ μA at 4.2 K, is due to Joule heating of the nanotube electron system.[13] The small peak in the resistance around zero bias current at $T_b = 4.2$ K, and the rise of $R(T)$ below ≈ 10 K, is attributed to non-ohmic contacts. At 77 K, and at 4.2 K for $|I| > 0.4$ μA, the dc contact resistance is ohmic and near the quantum value, $R_q \approx$ 6.4 kΩ. We thus expect that, in this regime of temperature and current, our radio frequency measurements correctly predict the relevant response at THz frequencies. This is not the case if the contact resistance is $>> R_q$.[4]

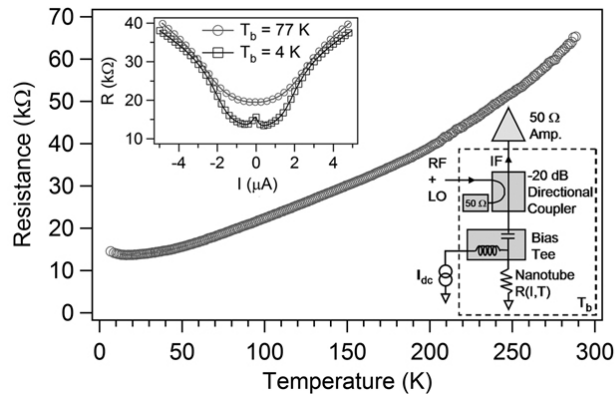

Figure 1. DC resistance $R = V/I$ as a function of bath temperature of 5 μm nanotube sample measured with a bias current of 100 nA. Top inset: Measured dc resistance as a function of bias current at bath temperatures of 4.2 K at 77 K. Bottom inset: Experimental schematic for rf heterodyne mixing measurement.

We characterize the sample using radio frequency (rf) heterodyne mixing. Two rf signals with frequencies $f_{rf} \approx$ 100 MHz are coupled to the nanotube through the coupled port of a directional coupler and the rf port of a bias tee, as shown in figure 1. The rf inputs have equal amplitude, resulting in 100% amplitude modulation of the input signal at the difference frequency, $f_{if} \approx$ 10 MHz. The nanotube is biased with a dc current through the dc port of the bias tee. The voltage change at $f_{if}$ is measured with a 50 Ω low-noise amplifier at the through port of the directional coupler. Hence, the nanotube sees a current bias (open circuit) at dc, while at $f_{rf}$ and $f_{if}$ it sees a 50 Ω load. We restrict the measurement to frequencies $f_{rf} \ll$ 1 GHz because the sample and on-chip wiring were not designed for higher frequency electrical coupling. With an appropriate substrate, gate, and electrical coupling structures, the same measurement could be performed at GHz or much higher frequencies.

The intrinsic voltage responsivity $S_V$ is defined as the change in the rms voltage across the device at $f_{if}$ divided by the change in rf power coupled to the device. In figure 2 we plot $S_V$ determined from measurement as a function of the dc bias current at $T_b =$ 4.2 K and 77 K. At $T_b =$ 4.2 K, the coupled rf input power was $\approx$ 10 nW. At $T_b =$ 77 K, it was increased to $\approx$ 100 nW because of the smaller response and reduced signal-to-noise ratio at 77 K. The rf input power coupled to the nanotube is what we use to compute $S_V$, and is significantly smaller than the available rf power (the power that would be coupled into a matched load) due to the high resistance of the nanotube. For $T_b =$ 4.2 K, we also plot the measured noise floor, which is not constant because the device resistance, and hence the coupling efficiency, changes with the bias current. At $T_b =$ 77 K, the noise floor is not shown, as it is below the measured data due to the larger signal power.

$S_V$ can be due either to bolometric detection or to a non-thermal *I-V* nonlinearity. In the limit where $f_{if}$ is small compared to the inverse of the thermal response time, the intrinsic voltage responsivity due to bolometric detection $S_{V,bolo}$ is given by[14]

$$S_{V,bolo} = \frac{\frac{I}{G}\left(\frac{dR}{dT}\right)}{1+\left(\frac{I^2}{G}\frac{dR}{dT}\right)\left(\frac{R-R_L}{R+R_L}\right)} \quad (1)$$

where *G* is the thermal conductance, *T* is the nanotube temperature, and $R_L$ is the load resistance seen by the nanotube at $f_{if}$, in this case 50 Ω. Electrothermal feedback is accounted for in the second term in the denominator, where a value of zero corresponds to no electrothermal feedback. We use *R(T)* measured with a small bias current and numerically differentiate to get *dR/dT*. Since *R* is between 10 and 40 kΩ, $R_L$ significantly loads down the voltage across the nanotube. It is this loaded-down IF voltage that we measure. We infer the *intrinsic* voltage responsivity using a voltage divider formula, to compare to equation 1.

In recent work, we used Johnson noise thermometry to determine the average electron temperature of a dc Joule-heated nanotube.[13] This enabled a direct determination of the thermal conductance for heat to escape the nanotube electron system as a function of the bias current, *G(I)*. Those measurements were performed on the same sample studied in the present work. We thus calculate $S_{V,bolo}$ from equation 1 using that thermal conductance. *R* and *dR/dT* are determined from the dc data of figure 1, and hence $S_{V,bolo}$ is predicted with no adjustable parameters. Previous studies of nanotube bundles treated the thermal conductance as a current-independent fitting parameter.[4] Our result for $S_{V,bolo}$ is plotted along with the measured data in figure 2. $S_{V,bolo}$ is not calculated near zero bias for $T_b = 4.2$ K because of the contact nonlinearity, discussed below.

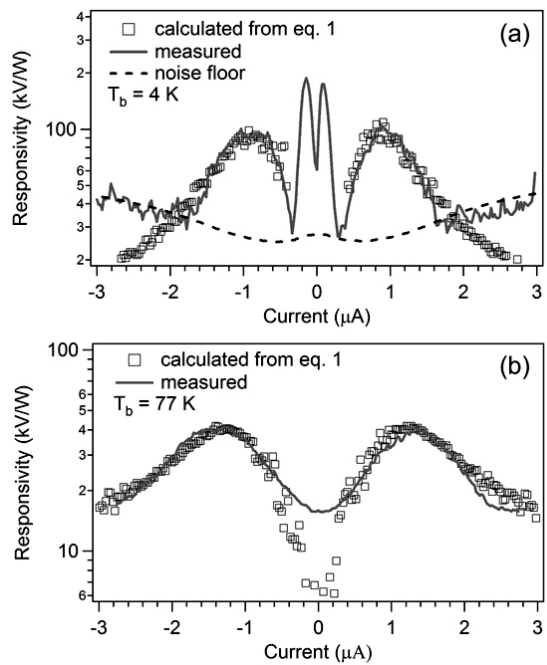

Figure 2. Intrinsic voltage responsivity $S_V$ from rf heterodyne mixing measurement, along with the bolometric responsivity calculated from equation 1. (a) Data at $T_b = 4.2$ K, along with the experimental noise floor. (b) Data at $T_b = 77$ K. The larger input power used at $T_b = 77$ K results in the discrepancy between the measured and calculated responsivities near zero bias current. The noise floor (not shown) is below the data.

At $T_b = 4.2$ K, we observe two distinct pairs of peaks. The outer peaks, at large bias current, are in good agreement with the bolometric responsivity calculated from equation 1 using the experimentally determined thermal conductance from ref. 13. The inner set of peaks is aligned with the low-bias feature in the $R(I)$ curve (figure 1), and we attribute these to the contact nonlinearity and not to a bolometric mechanism. At $T_b = 77$ K, only the outer set of peaks are seen. This response agrees well with the calculated bolometric responsivity except near zero bias current. The disagreement near zero bias current in figure 2(b) is due to the large rf input power used at $T_b = 77$ K. The oscillation at $f_{if}$ effectively averages over a range of bias currents, obscuring the dip predicted at zero bias current.

We find that the nanotube heterodyne response is explained by the bolometric mechanism at $T_b = 77$ and for $|I| > 0.4$ µA at $T_b = 4.2$ K. We now consider the response for $|I| < 0.4$ µA at $T_b = 4.2$ K. This response is aligned with the nonlinearity in the *I-V* curve due to non-ohmic contacts (figure 1). The heterodyne response is proportional to the second derivative of the dc *I-V* curve, provided that the *I-V* nonlinearity responds at the frequency of the applied rf voltage, and that the output load impedance for the dc *I-V* curve is the same as for the heterodyne response, or an appropriate correction factor can be applied. The voltage responsivity calculated from the nonlinear dc *I-V* curve, considering up to second order in the power series expansion of *V(I)*, is given by[15]

$$S_{V,non-lin.} = \frac{1}{2}\left(\frac{d^2V}{dI^2}\right)\frac{1}{R_{dyn}} \qquad (2)$$

where $R_{dyn} = dV/dI$ is the dynamic resistance. In figure 3 we plot $S_{V,non-lin.}$ calculated from equation 2 using the measured dc *I-V* curve at $T_b = 4.2$ K, along with the experimental $S_V$ for comparison. For $|I| < 0.4$ µA there is quantitative agreement between the measured result and the calculation from equation 2, consistent with a non-thermal response due to the contact *I-V* nonlinearity. The calculation from equation 2 displays a similar shape to the measured bolometric responsivity for $|I| > 0.4$ µA but does not display quantitative agreement. The *I-V* curve used to calculate equation 2 was measured with a dc current bias, which has a different load resistance (open circuit) than the load impedance at $f_{if}$ (50 Ω). This different load impedance affects the electrothermal feedback in the bolometric response (equation 1), resulting in the observed discrepancy. This different load impedance does not affect the response of the non-thermal contact nonlinearity (except for reducing the output coupling efficiency).

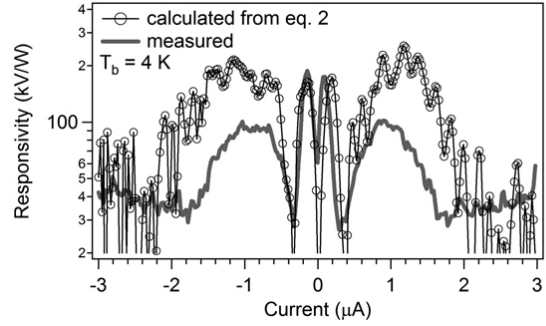

Figure 3. Responsivity calculated from the measured dc *I-V* curve (equation 2) as a function of bias current for the 5 μm nanotube at $T_b$ = 4.2 K. Also plotted is the measured responsivity.

Based on previous work, we expect the response due to the contact nonlinearity to be attenuated at THz frequencies.[4] The bolometric response should still apply in the THz region. We next consider the achievable IF bandwidth for bolometric detection. In the absence of electrothermal feedback, the thermal time constant $\tau_{th} = C/G$, where $C$ is the heat capacity. In the hot electron regime, the electrons act approximately as a separate thermal system from the nanotube phonons, and emitted phonons rapidly leave the nanotube. In this regime, $C$ is the electronic heat capacity. For $L$ = 5 μm, $T_b$ = 4.2 K, and $I$ = 1 μA, we expect $\tau_{th} \approx$ 4 ps.[13] This corresponds to an IF bandwidth $f_{3dB} = 1/(2\pi\tau_{th}) \approx$ 40 GHz. For a shorter nanotube, the IF bandwidth increases because of the added contribution to the thermal conductance from the out-diffusion of hot electrons into the contacts.[13] If instead the nanotube electron system is well coupled to the nanotube phonon system and the bottleneck for heat removal is the coupling of the nanotube phonon system to the environment, then the relevant heat capacity is the larger phonon heat capacity. In this case, for the same parameters as before, we expect $\tau_{th} \approx 200$ ps,[13] corresponding to $f_{3dB} \approx 1$ GHz. The electron-phonon decoupling needed to access the faster hot electron regime should be achievable at sufficiently low temperature. A direct measurement of the thermal time constant would clarify the limiting cooling mechanism, although such a measurement was not possible with the present sample.

For a nanotube THz detector, one would likely use antenna-coupling for the signal and LO, and an IF amplifier with a 50 Ω input impedance. This would have attendant coupling losses at the input and output due to the large nanotube resistance, ≈ 10-20 kΩ. It may be feasible to use a parallel array of nanotubes to significantly reduce the input and output coupling losses.[16] Even with the expected coupling loss, the required LO power is modest; ≈ 1 µW incident on the antenna is required to produce an LO current of 1 µA in the nanotube. By comparison, Schottky diodes require an LO power ~ mW.[9] Additionally, an ~µm long nanotube is predicted to exhibit plasmon standing wave resonances at THz frequencies,[17] although this has yet to be observed experimentally. Our simulations indicate that the input coupling efficiency would decrease, but only by approximately a factor of two, at the resonance peaks. If these challenges can be met, carbon nanotubes may prove to be an attractive high frequency detector technology, particularly for applications requiring a very large output bandwidth.

The work at Yale was supported by NSF grants DMR-0907082 and CHE-0911593. P. K. and M. S. P. acknowledge financial support from NSF NIRT (ECCS-0707748) and Honda R&D Co.